\documentclass[12pt,english]{article}
\usepackage[latin1]{inputenc}
\usepackage{color}
\makeatletter

\newcommand{\be}{\begin{equation}}
\newcommand{\ee}{\end{equation}}
\newcommand{\bea}{\begin{eqnarray}}
\newcommand{\eea}{\end{eqnarray}}
\newcommand{\ba}{\begin{array}}
\newcommand{\ea}{\end{array}}

\def\bbox{{\,\lower0.9pt\vbox{\hrule \hbox{\vrule height 0.2 cm
\hskip 0.2 cm \vrule height 0.2 cm}\hrule}\,}}
\newcommand{\dsl}{\pa \kern-0.5em /}

\def\tr{\rm tr}

\font\mybb=msbm10 at 10pt
\def\bb#1{\hbox{\mybb#1}}

\def\bR {\bb{R}}
\def\bE {\bb{E}}

\usepackage{babel}
\makeatother
\begin{document}


\begin{titlepage}
\rightline{DAMTP-06-24}
\rightline{\tt{hep-th/0604024}}

\vfill

\begin{center}
\baselineskip=16pt {\Large\bf  Self-gravitating Yang Monopoles}
\vskip 0.2cm
{\Large\bf in all Dimensions}
\vskip 0.3cm
{\large {\sl }}
\vskip 10.mm
{\bf G.W. Gibbons$^1$ and  P.K. Townsend$^2$}
\vskip 1cm
{\small
Department of Applied Mathematics and Theoretical Physics\\
Centre for Mathematical Sciences, University of Cambridge\\
Wilberforce Road, Cambridge, CB3 0WA, UK\\
}
\end{center}
\vfill

\par
\begin{center}
{\bf ABSTRACT}
\end{center}
\begin{quote}

The $(2k+2)$-dimensional Einstein-Yang-Mills equations for gauge group 
$SO(2k)$ (or $SU(2)$ for $k=2$ and $SU(3)$ for $k=3$) are shown 
to admit a family of spherically-symmetric  magnetic monopole
solutions, for both zero and non-zero cosmological constant $\Lambda$, 
characterized by a mass $m$ and a magnetic-type charge. The $k=1$
case is the Reissner-Nordstrom  black hole. The $k=2$ case yields a family of 
self-gravitating Yang monopoles. The asymptotic spacetime is Minkowski for 
$\Lambda=0$ and anti-de Sitter for $\Lambda<0$, but the total energy is infinite
for $k>1$. In all cases, there is an event horizon when $m>m_c$, for some 
critical mass $m_c$, which is {\it negative} for $k>1$. The horizon is degenerate 
when $m=m_c$, and the near-horizon solution is then an $adS_2\times
S^{2k}$ vacuum.

\vfill
\vfill
\vfill
\vfill
\vfill

\vfill
\hrule width 5.cm
\vskip 2.mm
{\small
\noindent $^1$  g.w.gibbons@damtp.cam.ac.uk\\
\noindent $^2$ p.k.townsend@damtp.cam.ac.uk
\\ }
\end{quote}
\end{titlepage}

\section{Introduction}
\setcounter{equation}{0}

In 1977 Yang found a spherically-symmetric  solution of the
5-dimensional Euclidean $SU(2)$ Yang-Mills (YM) equations that
provides a natural generalization of the Dirac magnetic monopole solution of 
Maxwell's equations in 3 space dimensions \cite{Yang:1977qv}. Just as
the Dirac monopole is a point  singularity in 3-space for which the 
integral of the magnetic 2-form is non-zero over any 
2-sphere enclosing the singularity, so the Yang monopole is a point 
singularity in 5-space for which the integral of $\tr (F\wedge F)$ is non-zero 
over any 4-sphere enclosing the singularity, where $F$ is the 
$su(2)$-valued YM 2-form field-strength. Just as the Dirac monopole 
is effectively a homogeneous and isotropic Maxwell field strength 
on the 2-sphere, so the Yang monopole is effectively 
a homogeneous and isotropic $SU(2)$ instanton on the 
4-sphere \cite{Jackiw:1976dw}. 
There are several important differences, however; for example, 
whereas Dirac's monopole charge can be any multiple of a 
fundamental unit, Yang's monopole 
charge has a fixed magnitude and only its sign can be freely 
chosen. In recent years, 
the Yang monopole has attracted  the attention of condensed matter 
theorists because of its relevance to the 4-dimensional Quantum Hall 
Effect \cite{Zhang:2001xs}. This application has led to the construction of 
some analogous higher-dimensional Yang-Monopole-type 
configurations \cite{Chen:1999ab,Fabinger:2002bk,
Bernevig:2003yz,Meng:2003vj,Hasebe:2003gx,Meng:2004qe}, 
which are effectively conformal maps to 
even-dimensional spheres of  higher-dimensional 
instantons \cite{Grossman:1984pi,Tchrakian:1978sf,Saclioglu:1986qn}. 

Neither the Yang monopole nor its higher-dimensional analogs have yet
had much impact on particle physics,  presumably because of the 
infra-red divergent magnetic field energy, as against the ultra-violet 
divergent energy of a Dirac monopole. An ultra-violet 
divergence can be removed by a cut-off that has no effect on
low-energy physics, but an infra-red divergence is more problematic.  
However,  infinite-energy `solitons'  are now commonplace in 
string/M-theory studies because there is often a physical
interpretation in terms of branes. In addition, it has recently been 
suggested that the endpoint of an $SO(32)$ heterotic string could be
viewed as a kind of higher-dimensional monopole
\cite{Polchinski:2005bg}. For the Yang monopole itself, the energy 
within a volume of radius $r$ diverges linearly with $r$, suggesting 
a possible interpretation as a `splayed' string. In confirmation of 
this interpretation, we note that the YM flux would get confined to 
an instanton flux tube in a Higgs phase, with a Yang-type monopole 
as its endpoint, just as the magnetic flux of  a Dirac magnetic
monopole would be confined to a magnetic flux tube in a 
superconducting vacuum.

These considerations suggest that the Yang monopole, and its 
higher-dimensional analogs, deserve further study. One purpose of 
this paper is to present a construction of Yang-type monopoles in 
$\bE^{2k+1}$ for gauge group $SO(2k)$. The construction is similar, 
but apparently not identical, to other constructions that have
appeared previously in the physics literature; from a purely mathematical
perspective, it has a straightforward generalization to  
homogeneous YM fields on homogeneous Einstein manifolds, but the
physical interpretation of such `generalized monopole' solutions
remains to be explored. For $k=1$ the construction 
yields the Dirac monopole and for $k=2$ it yields a superposition 
of Yang monopoles in each of the two $SU(2)$ factors of $SO(4)$; 
one may consistently truncate to a Yang-monopole solution of the 
$SU(2)$ YM  equations. For $k=3$ one may consistently truncate to
an $SU(3)$ subgroup of $SO(6)$. For $k=4$ the construction yields what 
could be called an ``octonionic monopole'' solution of the $SO(8)$
YM equations; this configuration was found in  \cite{Bernevig:2003yz} 
but it was not appreciated there that it solves  the YM equations.

All these solutions of the Euclidean YM equations can of course be 
viewed as static spherically-symmetric 
solutions of the YM equations on $(2k+2)$-dimensional 
Minkowski spacetime, and another purpose of this paper is to obtain 
the corresponding static solutions of the coupled Einstein-YM
equations. 
Our starting point for this purpose is the Lagrangian density
\be\label{lagstart}
{\cal L} = \sqrt{-\det g}\,  \left[ {1\over 16\pi G} \left(R -2\Lambda\right)
- {1\over2\gamma^2}\, \tr |F|^2  \right]\, ,
\ee
where $F$ is the field-strength for a YM theory with coupling constant 
$\gamma$, and $g$ is the metric of a $(2k+2)$-dimensional spacetime, 
and $\Lambda$ is the cosmological constant. Our principal interest is in the 
$k>1$ cases, and we find a self-gravitating Yang monopole 
for $k=2$, and investigate its global geometry. The $k>2$ 
self-gravitating ``Yang-type'' monopoles are 
found to have similar properties.
It turns out that  spherical symmetry allows a family of
self-gravitating solutions, parametrized by a mass $m$. For positive
$m$ there is an event horizon that `clothes' a spacelike singularity
and for $\Lambda=0$ one can view the resulting solution as a 
`Yang-monopole black hole' in the sense that the spacetime is asymptotically 
flat, albeit in a `weak' sense, since the total energy is infinite. 
For $\Lambda<0$ the solution is asymptotically anti-de Sitter (adS) in
the same sense.

For $\Lambda=0$ we present details of the global
structure as a function of $m$. A peculiarity (for $k>1$) is
that the absence of naked singularities is compatible with {\it
negative} $m$, for which the singularity is timelike. Specifically, 
there is a negative `critical' mass $m_c$ such that for $0>m>m_c$
there is both an inner (Cauchy) horizon and an outer (event) horizon;
the global structure in this case is similar to that of the
Reissner-Nordstrom (RN) black hole. When $m=m_c$, the two horizons
coincide to form a degenerate event horizon, like that of the extreme
RN black hole. Just as the near-horizon limit of  the extreme RN black
hole yields the Robinson-Bertotti $adS_2\times S^2$ vacuum of the 
4-dimensional Einstein-Maxwell theory, so the near-horizon limit of 
the extreme self-gravitating Yang-type monopole yields an  
$adS_2\times S^{2k}$ vacuum of the $(2k+2)$-dimensional 
Einstein-YM theory. For $k=2$, this type of vacuum has been 
discussed previously under the rubric of ``instanton-induced 
compactification''  \cite{Randjbar-Daemi:1983qa}, but the $k>2$ 
cases yield Kaluza-Klein vacua that, to our knowledge,  have not 
previously been considered. In any case, this type of vacuum may 
now be seen as a limit of  a more general solution, just  as many 
of the adS vacua of supergravity theories can be viewed as limits 
of brane spacetimes  \cite{Gibbons:1993sv}. 

For $\Lambda<0$, the global structure is more complicated but similar
for large adS radius. The spacetime is weakly asymptotically anti-de Sitter, 
because the metric tends to the adS metric but too slowly for the adS
energy to be finite. Nevertheless, the adS boundary is an $S^{2k}\times S^1$
spacetime, in which the $2k$-sphere has a background $\tr\, F^k$
magnetic field (or a uniform Maxwell magnetic 2-form $F$ in the 
$k=1$ case). But this is precisely, the vacuum of a quantum Hall 
system, so it is natural to conjecture that quantum gravity in a 
weakly asymptotically 
adS background  is dual to a noncommutative Quantum Hall theory. 
A boundary field theory at zero temperature would require a
$(2k+2)$-dimensional `bulk' solution with a degenerate event horizon, but 
there always exists such a solution. 

We should point out that self-gravitating Yang-type monopoles
differ in essential respects from the `monopole' solutions of the Einstein-YM 
equations found numerically by Bartnik and McKinnon
\cite{Bartnik:1988am}, and extended to the asymptotically
anti-de Sitter case and higher dimensions 
in \cite{Radu:2005mj}. In our notation, this  type of self-gravitating 
`monopole' would be 
a solution of $SO(2k+1)$ YM equations in a $(2k+2)$-dimensional 
spacetime, for any integer $2k$. These solutions are such that
the field strength has components in the radial direction, as well 
as on the $2k$-sphere. The absence of such radial components  is
presumably the difference that makes the self-gravitating Yang-type 
monopole equations tractable analytically.

 \section{Preliminaries}

The Euler-Lagrange equations for the Lagrangian density 
(\ref{lagstart}) consist of the sourceless YM 
equations in a general $(2k+2)$-dimensional spacetime, and 
Einstein equations for the metric of this spacetime with a 
YM source. The YM potential 1-form can be written in coordinates 
$\{x^m; m=0,1,\dots 2k+1\}$ as
\be
A= dx^m A_m^a \, T_a\, , 
\ee
where the matrices $T_a$ span a representation $R$ of the Lie 
algebra of the  (semi-simple) gauge group $G$. If $f_{ab}{}^c$ 
are the structure constants of the algebra then 
\be
[T_a,T_b] = if_{ab}{}^cT_c\, .  
\ee
We assume that $R$ is the fundamental representation and adopt 
the normalization convention
\be\label{tracenorm}
2\tr \, \left(T_a T_b\right) = \delta_{ab}\,  . 
\ee
In particular, for $G=SU(2)$ we have $T_a = \sigma_a/2$, where 
$\sigma_a$ ($a=1,2,3$) are the Pauli matrices. The YM equation is 
\be
\partial_m \sqrt{-\det g}\, F^{mn} = i \sqrt{-\det g}\, [A_m,F^{mn}] \, , 
\ee

The Einstein equation is
\be
G_{mn} = (8\pi G)T_{mn}  - g_{mn}\Lambda \, , 
\ee
where the YM stress tensor is
\be
T_{mn} = \gamma^{-2}\left[\tr \left(F_m{}^pF_{np}\right) - 
{1\over4}g_{mn}\tr \left(F_{pq}F^{pq}\right)\right]\, . 
\ee
For notational purposes it is convenient to consider the corresponding 
quadratic differential $T \equiv T_{mn}\, dx^mdx^n$.

We seek spherically-symmetric solutions of these equations for which 
the spacetime metric takes the form
\be\label{Scoord}
ds^2 = -\Delta\, dt^2  + \Delta^{-1} dr^2 + r^2 d\Omega_{2k}^2
\ee
for some function $\Delta(r)$, where $d\Omega_{2k}^2$ is the 
$SO(2k-1)$-invariant metric on the unit $2k$-sphere. These 
Schwarzschild-type coordinates are useful for many purposes, 
but it simpler to present the solution for the YM potential in 
terms of isotropic coordinates for which
\be\label{isotropic1}
ds^2 = -f dt^2 + \omega^2\left[d\rho^2 + \rho^2 d\Omega_{2k}^2\right]
\ee
for functions 
\be
f(\rho) = \Delta\left(r(\rho)\right)\, ,
\qquad \omega(\rho) = r(\rho)/\rho\, , 
\ee
where $r(\rho)$ is found by solving the differential equation 
\be\label{rtorho}
{dr\over r\sqrt{\Delta(r)}} = {d\rho\over\rho}\, . 
\ee 
We shall find it  convenient to introduce coordinates 
$\{n^i; i=1,\dots , 2k\}$ for $S^{2k}$ such that
\be\label{spacemetric}
d\Omega_{2k}^2  =  h_{ij} dn^idn^j 
\ee
where 
\be
h_{ij}=\delta_{ij} + {n^in^j\over 1-n^2}\, , 
\qquad \left(n^2 = \delta_{ij} n^in^j\right). 
\ee
Note that these coordinates cover only one hemi-$2k$-sphere and 
are singular on the equatorial $(2k-1)$-sphere at $n^2=1$.  The volume 
form on $S^{2k}$ in these coordinates is
\be
vol(S^{2k}) = (1-n^2)^{-{1\over2}} dn^1 \wedge \dots \wedge dn^{2k}\, . 
\ee

\section{The Yang monopole}

We now review the Yang-monopole solution for $k=2$  in a way 
that will also allow preliminary
consideration of the $k>2$ case, which we take up in the following section.  
The Yang monopole has 1-form potential 
\be\label{pot}
A= {\Sigma_{ij}\, n^idn^j \over 1+ \sqrt{1-n^2}} \, , 
\ee
where
\be 
\Sigma_{ij} = \eta_{ij}^a T_a
\ee
for constants $\eta^a_{ij}$ satisfying
\be\label{thooft1}
\eta_{ij}^a\eta_{kl}^b f_{ab}{}^c = 2\left( \delta_{l[i}\eta_{j]k}^c 
- \delta_{k[i}\eta_{j]l}^c\right)\, . 
\ee
Using this relation, it can be shown that the only non-vanishing 
components of $F$ are
\be
F_{ij} = \Sigma_{ij}  - 2{\delta_{l[i}\Sigma_{j]k} n^ln^k \over 1-n^2 
+ \sqrt{1-n^2}}\, . 
\ee
We also record here that
\be
F^{ij} \equiv \rho^{-4} h^{ik}h^{jl} F_{kl} = 
\rho^{-4}\left(\Sigma_{ij} + 2 {\delta_{l[i}\Sigma_{j]k}^a n^ln^k\over 1+ 
\sqrt{1-n^2}}\right)\, . 
\ee
Using (\ref{thooft1}) again, it may be verified that
\be\label{eqcalc}
( \sqrt{\det h})^{-1}\partial_i \left(\sqrt{\det h}F^{ij}\right) 
= -i [A_i,F^{ij}] 
= {2(k-1)   \over (1 + \sqrt{1-n^2})\rho^4}\, n_k \,\Sigma_{jk}\, , 
\ee
and hence that the YM equations are satisfied. This 
result is conditional on the existence of an 
constant ``$\eta$-tensor'' satisfying (\ref{thooft1}) 
but this {\it is} satisfied for $k=2$ and gauge group 
$SU(2)$ by the self-dual 't Hooft tensor (a quaternionic 
structure on $\bE^4$), and this fact 
motivates the notation. Note that the anti-self-dual 't Hooft 
tensor $\bar\eta_{ij}^a$, also satisfies
(\ref{thooft1}) and this yields a Yang anti-monopole. As shown 
by Yang \cite{Yang:1977qv}, 
these are the only possibilities compatible with spherical symmetry
(which is not manifest  
in the above formula for $F$ but which will be seen to be a property
of all gauge-invariant 
quantities constructed from $F$). 

It can be shown that
\be
\tr\,  F^k =  \left(1-n^2\right)^{-{1\over2}}\, \tr \, \Sigma^k\,  
\qquad  \left(\Sigma = {1\over2}\Sigma_{ij}\,  dn^i  dn^j \right)\, ,
\ee
where the exterior product of forms is implicit.
It follows that $\tr\, F^k$ is proportional to the volume form on $S^{2k}$.
For $k=2$ we may use (\ref{tracenorm}), and the 't Hooft tensor identities
\be\label{thooft2}
\eta_{[ij}^a \eta_{kl]}^a  \equiv \varepsilon_{ijkl}\, , \qquad
\bar\eta_{[ij}^a \bar\eta_{kl]}^a  \equiv -\varepsilon_{ijkl}\, ,
\ee
to show that
\be
\pm {1\over 3} \tr \, F^2 = \left(1-n^2\right)^{-{1\over2}}\, 
dn^1\wedge dn^2\wedge dn^3\wedge dn^4\, , 
\ee
which is the volume form on the unit 4-sphere. As the unit 4-sphere 
has 4-volume 
$8\pi^2/3$, we deduce that
\be
{1\over 8\pi^2}\int_{S^4} \tr\, F\wedge F =\pm 1\, , 
\ee
and hence that the Yang monopole configuration is equivalent to an 
instanton or anti-instanton on a 4-sphere  \cite{Jackiw:1976dw}. 

To conclude our discussion of the flat-space Yang monopole, we note that if
\be\label{thooft3}
\tr \left(\Sigma_{ik}\Sigma_{kj}\right) = -{1\over2} N \delta_{ij}\, , 
\ee
where the constant $N$ must be such that
\be
\sum_{i<j} \tr\, \Sigma_{ij} \Sigma_{ij} = {kN\over2}\, , 
\ee
then
\be
\tr \left(F_{ik}F^{jk} \right) = {N\over 2\rho^4} \,  \delta_i^j\, . 
\ee
The  energy density is therefore
\be
{\cal E}(\rho) \equiv T_{00} = {1\over 4\gamma^2}\tr 
\left(F_{ij}F^{ij}\right) = 
{kN\over 4\gamma^2\rho^4}\,   ,  
\ee
which is spherically-symmetric. It follows (for $k>1$) that the 
total energy within a radius $R$ grows like 
$R^{2k-3}$. For $k=2$, and $G=SU(2)$, the `$\eta$-tensor' is 
the 't Hooft tensor and (\ref{thooft3}) is satisfied with $N=3$.
The total energy of the Yang-monopole within a radius $R$ is therefore 
\be
E(R) = {8\pi^2\over3}\int_0^R\! {\cal E} \, d\rho = (2\pi/\gamma)^2 R\, . 
\ee
The coefficient $(2\pi/\gamma)^2$ is precisely the tension of the instanton string, 
so the Yang monopole is marginally stable against collapse to an instanton flux string,
although (as mentioned in the introduction) it will likely be unstable if the $SU(2)$ gauge 
symmetry is spontaneously broken.

\section{Higher-dimensional monopoles}

For gauge group $SO(2k)$ the  `$\eta$-tensor'  relation
(\ref{thooft1}) 
is equivalent  to 
the $so(2k)$ commutation relations
\be\label{comrel}
[\Sigma_{ij},\Sigma_{kl}] = 2i\left(\delta_{l[i}\Sigma_{j]k} 
- \delta_{k[i}\Sigma_{j]l}\right)\, . 
\ee
The YM equations are therefore satisfied by $A$ of the form 
(\ref{pot}) provided that  the $k(2k-1)$ matrices $\Sigma_{ij}$ 
span the Lie algebra $so(2k)$. In this case the constants 
$\eta_{ij}^a$ just specify the linear transformation 
from the basis of $so(2k)$ provided by the matrices $T_a$ to the 
basis provided by the matrices $\Sigma_{ij}$. Let $\omega_{ij}$ 
be the components of $A$ in this new basis:
\be
A = {1\over2} \omega_{ij} \Sigma_{ij}\, . 
\ee
We can express these components in dyad form as
\be\label{dyadform}
\omega = {\tilde {\bf n}\, d{\bf n} - d\tilde{\bf n}\,  
{\bf n}\over 1+ \sqrt{1-n^2}}\, , 
\ee
where ${\bf n}$ is the column $2k$-vector with components $n^i$, 
and $\tilde{\bf n}$ its transposed row vector.  This $SO(2k)$ 
gauge potential satisfies the YM equations on $S^{2k}$ and hence 
the YM equations  on $\bE^{2k+1}$. In fact, this applies not only 
for integer $k$ but also for half-integer $k$, in the sense that 
an SO(n) potential of the form (\ref{pot}) solves the YM equation 
on $\bE^{n+1}$ for all $n\ge2$. 

This result has a simple geometrical interpretation, which we now 
present. Recall that the 
potential of a Dirac monopole is the connection on the principal 
$SO(2)$ bundle over a unit 2-sphere  that has a 3-sphere 
as the total space. The metric on this 3-sphere takes the form  
$(\sigma -\omega)^2 + d\Omega_2^2$, 
where $\sigma=d\varphi$ is the invariant 1-form on the $S^1$ fibre, 
in its parametrization by the angle $\varphi$, and $d\Omega_2^2$ 
is the $SO(3)$-invariant metric on the 2-sphere base. The $SO(2)$
connection $\omega$ needed such that the bundle metric is that of 
the unit 3-sphere is precisely  the $n=2$ ($k=1$) case of 
(\ref{dyadform}).  In general, the potential (\ref{dyadform}) 
is the connection 
for the principal  $SO(2k)$ bundle over a unit $2k$-sphere that 
has $SO(2k+1)$ as its total space. 
To see this\footnote{Here we follow Appendix B of
  \cite{Cvetic:2003jy}, 
where a Kaluza-Klein interpretation can be found.}, we express 
$g \in SO(n+1)$ as $g=hk$, where $h\in SO(n)$  and $k$ 
is a representative of the coset space $SO(n+1)/SO(n)\cong S^n$. 
The bi-invariant 
metric on $SO(n+1)$ is
\be
dl^2 = {1 \over 2} \tr \left( g^{-1} dg \right)^2=  
{1 \over 2} \tr \left( h^{-1} dh + dk k^{-1} \right)^2\, , 
\ee
where the trace is over the $(n+1)\times(n+1)$ matrices of the 
vector representation. To 
express this metric in the required form, we need expressions 
for the left-invariant forms 
$h^{-1}dh$ on $SO(n)$ and the right-invariant forms $dk\, k^{-1}$ 
on $S^n$. For the obvious 
embedding of $SO(n)$ in $SO(n+1)$ we have
\be
h^{-1}dh = \pmatrix{ \sigma &0 \cr 0 &0}\, ,
\ee
for left-invariant forms $\sigma$ on $SO(n)$. For $k$ we may choose
\be
k= \exp \thinspace {\pmatrix { 0 & {\bf b}\cr -\tilde{\bf b} & 0 \cr }}
= \pmatrix { 1-{{\bf n} \tilde{\bf n} \over 1+ \sqrt{1-n^2}} & {\bf n}
 \cr -\tilde{\bf n} & \sqrt{1-n^2}\cr }\, , 
\ee
where
\be
{\bf n} = (\sin |{\bf b}| / |{\bf b}|)\  {\bf b}\, . 
\ee
A calculation then shows that
\be
k^{-1} dk=   \pmatrix{ \omega & {\bf v} \cr 
- \tilde{\bf v} & 0}\, , \qquad
dk \, k^{-1}=   \pmatrix{ -\omega & {\bf v} \cr 
- \tilde{\bf v} & 0}\, ,
\ee
where $\omega$ is precisely the connection of (\ref{dyadform}), and 
\be
{\bf v} = d{\bf n} +  {\bf n} \, {(\tilde{\bf n}d{\bf n})\over 
1 -n^2 + \sqrt{1-n^2}} \, . 
\ee
Thus,
\be
h^{-1} dh + dk k^{-1}  = \pmatrix{\sigma- \omega & {\bf v} \cr 
-\tilde{\bf v}&0}\, , 
\ee
and hence the bi-invariant metric on $SO(n+1)$ is
\be
dl^2 = {1\over 2}\tr\left(\sigma  -\omega \right)^2  + 
\tilde{\bf v}{\bf v} \, ,  
\ee
where the trace is now over the $n\times n$ matrices of 
the vector representation of $SO(n)$. 
One finds that
\be
\tilde{\bf v}{\bf v}  = d\tilde{\bf n}
\left(1 - {{\bf n}\tilde{\bf n} \over 1-n^2}\right) d{\bf n}\, ,
\ee
which is the metric on the unit  $n$-sphere.

Although this construction applies for any $n$, we must distinguish between 
even $n$ and odd $n$. For $n=2k$ with integer $k$
the YM configuration on $S^{2k}$ is topologically non-trivial because
the integral of $\tr\, F^k$ over the $2k$-sphere is non-zero. 
This integral is a
conserved magnetic charge that justifies the `monopole' terminology. As we have
seen, the $k=1$ case yields the Dirac monopole. The $k=2$ case yields 
a solution of the $SO(4)$ YM equations that is a 
superposition of Yang monopoles in each of the two 
$SU(2)$ factors of $SO(4)$. For $k=4$ we have an ``octonionic
monopole'' solution of the $SO(8)$ YM equations. 
In contrast, the odd-sphere cases are better thought of as
$SO(n+1)$-invariant instantons on $\bE^{n+1}$. For $n=3$, the
construction yields an $SO(4)$-invariant configuration of $SO(3)$ 
YM fields on the 3-sphere, but this is just the spherically-symmetric 
1-instanton solution on $\bE^4$. Similarly, for $n=5$ we find an
$SO(6)$-invariant instanton of the $SO(5)$ YM equations on $\bE^6$, 
and for $n=7$ an $SO(8)$-invariant instanton of the $SO(7)$ YM
equations on $\bE^8$. The latter is presumably related to the
``octonionic instanton'' of \cite{Fairlie:1984mp,Fubini:1985jm}.

\section{Self-gravitating Yang-type monopoles}

Because a Yang-type monopole has a field-strength 2-form with components 
only on the $2k$-sphere, it can be viewed as a static solution in 
any  $(2k+2)$-dimensional 
spacetime with metric of the form (\ref{isotropic1}). However, we 
then need to solve the
Einstein equation for the YM source to determine the functions 
appearing in this metric 
ansatz. A Yang-type monopole has the spherically-symmetric 
(quadratic differential) stress tensor 
\be
T= {kN\over 4\gamma^2\omega^4\rho^4} \left[ f dt^2 + 
\omega^2 d\rho^2 - [(k-2)/k]\omega^2 \rho^2 d\Omega_{2k}^2\right]\, , 
\ee
where $N$ is the constant appearing in (\ref{thooft3}). 
In the Schwarzschild-type coordinates, for which the spacetime 
metric is (\ref{Scoord}), we have
\be
T= {kN\over 4\gamma^2r^4} \left[ \Delta \, dt^2 + \Delta^{-1} dr^2 -  
[(k-2)/k]r^2 d\Omega_{2k}^2\right]\, .
\ee

If we write
\be
\Delta (r) = 1- {2GM(r)\over r^{2k-1}}\, ,  
\ee 
for ``mass function'' $M(r)$, then the Einstein equations reduce, 
for $\Lambda=0$,  to the ordinary differential equation\footnote{See,
  for example, section 6 of \cite{Gibbons:2005vp}.}
\be
{dM\over dr} = {N\pi \over  \gamma^2}\,\, r^{2(k-2)} + {\Lambda\over
  2kG}\,  r^{2k}\, . 
\ee
The solution is\footnote{This also applies for half-integer $k$, with
  the exception of $k=3/2$. The $k=3/2$ case was analysed 
in \cite{Okuyama:2002mh}.}
\be
M(r) = m + {N\pi\over (2k-3)\gamma^2}\, r^{2k-3} + {(\Lambda/G)\over
  2k(2k+1)} \, r^{2k+1}\, , 
\ee
for arbitrary constant $m$, which has dimensions of mass. This yields
\be
\Delta = 1 -{2Gm\over r^{2k-1}} - {\mu^2\over r^2} - 
{\Lambda r^2\over k(2k+1)}\, , 
\ee
where 
\be
\mu^2 = {2\pi G N \over (2k-3)\gamma^2}\, . 
\ee
Note that $\mu^2<0$ for $k=1$, in which case we should set $N=1$; for
$\Lambda=0$ we then have the Reissner-Nordstrom metric. 
We now consider in turn the subcases of zero, negative and positive
$\Lambda$. 

\subsection{$\Lambda$=0} 

For $\Lambda=0$, the spacetime metric is 
\be
ds^2 = - {P(r) \over r^{2k-1}}dt^2 + {r^{2k-1}\over P(r)} dr^2 + 
r^2d\Omega_{2k}^2\, , 
\ee
where $P(r)$ is the polynomial
\be
P(r) = r^{2k-1} - \mu^2 r^{2k-3} -2Gm\, . 
\ee

When $m=0$ we have a `pure' self-gravitating Yang-type monopole 
with spacetime metric
\be
ds^2 = -\left(1-{\mu^2\over r^2}\right)dt^2 + 
\left(1-{\mu^2\over r^2}\right)^{-1} dr^2 + 
r^2d\Omega_{2k}^2\, . 
\ee
There is an event horizon at $r=\mu$ behind which we find a 
spacelike singularity at $r=0$. This is therefore a kind of 
Yang-monopole black hole. Integration of (\ref{rtorho}) 
with $rH(r)=\sqrt{r^2-\mu^2}$ yields
\be
r= {\mu\over 2}\left({\rho\over \rho_0} + {\rho_0\over\rho}\right)\, , 
\ee
for arbitrary constant $\rho_0$. We see that isotropic coordinates 
cover only the region outside the horizon, as expected, but also 
that there are two isometric exterior regions, with $\rho>\rho_0$ 
and $\rho<\rho_0$. These regions are connected on a hypersurface 
of constant $t$ by  a  `throat'  with minimum  $2k$-sphere of 
radius $\mu$ at $\rho=\rho_0$. The $\tr\, F^k$ flux of the monopole 
passes through this throat. Note that in the region $\rho < \rho_0$, 
the singularity of the YM field at $\rho=0$ corresponds to a source 
of the flux at $r=\infty$. 

For `impure' solutions with $m\ne0$ the horizon structure may be 
more complex. Note that the polynomial $P(r)$ has a minimum at 
positive $r$ and a maximum at negative $r$, and {\it no other
  extrema}. It follows that  $P$ has at most three real roots. Let 
us consider how these roots change with $m$, starting with  $m=0$, 
where we have one real postive root, one real negative root and 
one root at $r=0$. As $m$ increases from zero, the root at $r=0$ 
moves to negative $r$, where it must either stay as $m$  continues 
to increase monotonically (since a root can cross the $r=0$ axis 
only at $m=0$) or
disappear (by coalescence with the other negative root, which must 
also stay negative). Thus, 
there is a unique real positive root $r=r_+$ for $m>0$, 
corresponding to a non-degenerate
event horizon behind which there is a spacelike singularity 
at $r=0$. Moreover, $r_+$ increases monotonically with $m$, so 
that matter falling into the `black hole' increases the surface 
area of the horizon, as one would expect. 

However, $m$ need not be positive. As we decrease $m$ from zero, 
the zero root of $P(r)$ now moves to positive $r$, so we now have 
two real positive zeros of $P(r)$, corresponding to an inner 
(Cauchy) horizon at $r=r_-$ and an outer (event) horizon at 
$r=r_+>r_-$. As before, this state of affairs cannot change 
as $m$ continues to decrease monotonically to ever larger 
negative values until the two roots 
$r=r_\pm$ coincide, after which there is no longer any real 
positive root and hence no event horizon. 
Thus, there exists a `critical' {\it negative} value $m_c$ of 
$m$, such that there is an event horizon for $m>m_c$ but a 
naked (timelike) singularity for $m<m_c$. In  fact, 
\be
Gm_c =  - {1\over 2k-3} \left({2\pi GN \over (2k-1)
\gamma^2}\right)^{k-{1\over2}}\, . 
\ee
{}For $m_c<m<0$ the global structure is similar to that of the 
Reissner-Nordstrom black hole. When $m=m_c$, the inner and 
outer horizons coalesce to form a degenerate (zero surface 
gravity) event horizon. The global geometry is then similar 
to that of the extreme Reissner-Nordstrom spacetime,
and the  near-horizon limit of this  `extremal' 
self-gravitating Yang-type monopole is an $adS_2\times S^{2k}$ 
vacuum solution. 

To illustrate these general observations, we will consider the 
$k=2$ (Yang-monopole) case in more detail. In this case the 
polynomial $P(r)$ is cubic, 
\be
P(r) = r^3 -\mu^2 r -2Gm\, , 
\ee
and there are no naked singularities as long as 
$3\sqrt{3} \, Gm/\mu^3 \ge -1$, or
\be
m\gamma^3/\sqrt{G} \ge -(2\pi)^{3\over2}\, . 
\ee
At saturation, the cubic polynomial is
 \be
P(r)= \left(r+ {2\mu\over \sqrt{3}}\right)
\left(r- {\mu\over\sqrt{3}}\right)^2\, .
\ee
The double zero at $r= \mu/\sqrt{3}$ means that the inner and 
outer horizons have 
now merged to form a single degenerate horizon at $r= \mu/\sqrt{3}$. Since
\be
{d\rho\over \rho} = {dr\over \left(r-{\mu\over \sqrt{3}}\right) 
\sqrt{1+ {2\mu \over \sqrt{3}\, r}}}\, , 
\ee
we have, near the horizon at $r= \mu/\sqrt{3}$,
\be
\rho^{\sqrt{3}} \sim \left(r-{\mu\over \sqrt{3}}\right)
\ee
so that $\rho \to 0$ at the horizon. Introducing the new time variable
\be
\lambda = {3\sqrt{3}\over \mu^2} t \, , 
\ee
we may write the near horizon metric as
\be
ds^2=  {\mu^2\over 3}\left[ -\rho^{2\sqrt{3}} d\lambda^2 +  
\left({d\rho\over\rho}\right)^2
+ d\Omega_4^2\right] .
\ee
This is a metric for $adS_2\times S^4$, and the singularity 
at $\rho=0$ is a coordinate 
singularity at the adS horizon. 

\subsection{$\Lambda<0$}

For a `pure' $m=0$ monopole, we have  
\be
\Delta = 1 - {\mu^2\over r^2} + {r^2\over \ell^2}
\ee
for $\Lambda<0$, where 
\be
\ell = \sqrt{k(2k+1)|/|\Lambda|}
\ee
is the adS radius. There is a unique non-degenerate horizon at
\be
r = {1\over \sqrt{2}}\sqrt{\sqrt{\ell^4 + 4\mu^2\ell^2} - \ell^2} \, . 
\ee

For simplicity, we shall concentrate for $m\ne0$ on the $k=2$, Yang-monopole,
case, for which 
\be
\Delta = 1 - {2Gm\over r^3} -{\mu^2\over r^2} + {r^2\over \ell^2}\, .  
\ee
The horizon structure is qualitatively similar for higher $k$.
For positive $m$ there is still a unique non-degenerate 
event horizon. For negative $m$ there is
a critical mass $m_c$ such that there exists both an inner and outer
horizon for $0>m>m_c$, with a unique degenerate horizon when $m=m_c$,
and a naked singularity when $m<m_c$. For $k=2$ we have 
\be 
m_c = - r_c^2\left(1+ 2r_c^2/\ell^2\right)\, , 
\ee
where $r_c$ is the position of the degenerate horizon, given by
\be
10 r_c^2 = \ell \left(\sqrt{9\ell^2 + 20 \mu^2} - 
3\ell\right)\, . 
\ee
The global structure is thus similar to the $\Lambda=0$ case apart
from the asymptotic behaviour for large $r$. 

\subsection{$\Lambda>0$}

For a `pure' $m=0$ monopole we now have
\be
\Delta = 1-{\mu^2\over r^2} - {r^2\over R^2}
\ee
where 
\be
R = \sqrt{k(2k+1)|/\Lambda}
\ee
is the de-Sitter radius (the radius of the minimal $2k$-sphere).
Now there is both an inner (event) and outer (cosmological) horizon 
for $R>2\mu$. These horizons coalesce at $R=2\mu$, and for $R<2\mu$ 
there is a naked singularity at $r=0$. 

For $m\ne0$ we shall again concentrate on the $k=2$ case, for which
\be
\Delta  = 1-{2Gm\over r^3} -{\mu^2\over r^2} - {r^2\over R^2}\, .
\ee
Let us consider increasing $m$ monotonically from zero. If there 
were initially two zeros of $\Delta$ then this will continue to be the 
case up to a maximum value of $m$, after which there is only a 
naked singularity. For $m<0$ there may be three horizons, 
an inner (Cauchy) horizon, an outer (event) horizon and a cosmological 
horizon, provided that
\be
R> R_c \equiv {2\sqrt{5}\over 3}\, \mu\, , 
\ee
and in this case there will be some negative $m$ for which 
the inner and outer horizons coalesce, yielding just two horizons, 
one a degenerate event horizon and the other the cosmological horizon.
For $R\le R_c$ there is only a cosmological horizon.

\section{Discussion}

We have shown that the interpretation of $SO(2k+1)$ as an $SO(2k)$
bundle over $S^{2k}$ leads to a non-singular $SO(2k+1)$-invariant
configuration of $SO(2k)$ Yang-Mills fields on $S^{2k}$ with
topological charge proportional to the integral of $\tr\, F^k$ over the
$2k$-sphere. Viewed as a singular configuration on $\bE^{2k+1}$,
these are magnetic monopoles: $k=1$ yields the Dirac monopole and 
$k=2$ the Yang monopole; more precisely, a superposition of two Yang 
monopoles that is trivially truncated to a single Yang monopole. 
The $k>2$ cases yield higher-dimensional magnetic monopoles, such as 
a $k=4$ ``octonionic'' monopole. 

The ``Yang-type'' monopoles can be viewed as static
spherically-symmetric solutions of the $SO(2k)$ YM equations 
on $(2k+2)$-dimensional Minkowski space.  
We have extended these solutions of the flat-space YM equations 
to a class of static spherically-symmetric solutions of the
$(2k+2)$-dimensional Einstein-Yang-Mills equations,  
parametrized by a mass $m$, for arbitrary cosmological constant
$\Lambda$. 

When $\Lambda=0$, these solutions generalise the 
magnetic Reissner-Nordstrom black hole, which is the $k=1$ case. 
The $k=2$ case is a self-gravitating version of the Yang magnetic monopole, 
and there is an analogous solution for all higher $k$. All $k>1$
solutions have infinite total energy, the energy within a 
radius $r$ diverging as $r^{2k-3}$. In all cases, there is an  
event horizon for $m>m_c$ where the `critical'  value $m_c$ of $m$ is
{\it negative} (for $k\ge2$). For $m_c<m<0$ the global structure 
is similar to that of the Reissner-Nordstrom black hole, and the 
$m=m_c$ solution is similar to the extreme Reissner-Nordstrom 
black hole. Specifically, the horizon is degenerate when $m=m_c$ 
and the near-horizon solution is an $adS_2\times S^{2k}$ vacuum. 
For $k=1$ this is just the Robinson-Bertotti solution 
of the Einstein-Maxwell equations and for $k=2$ it yields the 
solutions found previously by ``instanton-induced 
compactification'' \cite{Randjbar-Daemi:1983qa}.  
The $k>2$ cases yield new Einstein-YM vacua. 

The similarity of the extremal self-gravitating monopoles 
 to the extreme Reissner-Nordstrom black hole suggests a connection with 
supersymmetry. However, the total mass of either the Yang
monopole or its higher-dimensional analogs is undefined, so 
it is hard to see how any well-defined charges given by integrals 
at spatial infinity could span a supersymmetry algebra.  It is 
also difficult to see how Yang-type monopole solutions could be 
extended to solutions of supergravity theories. For example, 
6-dimensional supergravity-YM theories have a 3-form field 
strength $H$ that obeys an `anomalous' Bianchi identity of the 
form $dH\propto \tr F^2$. It follows that $H$ cannot vanish in 
a Yang monopole background, but no non-zero $H$ could be 
compatible with spherical symmetry. 

For $\Lambda<0$ one finds analogous `monopole black hole'  solutions
for which the spacetime metric is  weakly asymptotically adS . 
It is natural to suppose 
that the adS boundary is the vacuum for some holographically 
dual field theory, which 
must be a non-commutative Quantum Hall type theory as a consequence of the 
`magnetic' flux through the $2k$-sphere  of the $S^{2k}\times S^1$ boundary. 
Thus, the results presented here potentially provide a new 
way to study non-commutative
field theories as holographic duals of gravitational theories. 

Another application of our results is to wormhole solutions 
of the Euclidean Einstein-YM equations on $\bR\times S^n$. For 
$\Lambda=0$, the metric is
\be
ds^2 = d\tau^2 + a^2(\tau)d\Omega_n^2
\ee
where $a^2(\tau)= \tau^2 + r_0^2$ for constant $r_0$. One may also
consider Euclidean solutions of this type for $\Lambda\ne0$. 
The construction for general $n$ is a straightforward extension of 
the $n=3$ case discussed in \cite{Hosoya:1989zn}. 

We conclude with a comment on the $k=3$ case.
Our construction of higher-dimensional monopoles rests on the 
interpretation of $S^{2k}$ as the homogeneous space $SO(2k+1)/SO(2k)$,
which is unique (at least locally) for all integer $k$ except $k=3$.
The group $G_2$ acts transitively on $S^6$, leaving invariant the
round metric (see e.g. \cite{Macfarlane:2002hr}). The stabilizer 
is $SU(3)$, so 
\be
S^6 \cong G_2/SU(3)\, ,   
\ee
and this implies the existence of a $G_2$-invariant monopole solution 
of the $SU(3)$ YM equations on $\bE^7$. The $G_2$-symmetry of this
solution {\it is} compatible with non-zero $H$, and so a
self-gravitating version of it might be extendible to a
membrane-monopole of the heterotic string (not to be confused 
with the ``octonionic membrane'' of \cite{Ivanova:1993nu,Gunaydin:1995ku},
which is a solution of the equations of 10-dimensional supergravity
coupled to a YM supermultiplet with $G_2$ gauge group).
We leave this to a future investigation.

\noindent

\vskip 1cm
\noindent
{\bf Acknowledgements:} 
 PKT is supported by a Senior  Research Fellowship of the EPSRC. 
We thank  Eric Bergshoeff and Harvey Reall for discussions.



\begin{thebibliography}{99}

\bibitem{Yang:1977qv}
C.~N.~Yang,
{\it Generalization Of Dirac's Monopole To SU(2) Gauge Fields}, 
J.\ Math.\ Phys.\  {\bf 19}, 320 (1978).

\bibitem{Jackiw:1976dw}
R.~Jackiw and C.~Rebbi,
{\it Conformal Properties Of A Yang-Mills Pseudoparticle}, 
Phys.\ Rev.\ D {\bf 14}, 517 (1976).

\bibitem{Zhang:2001xs}
  S.~C.~Zhang and J.~p.~Hu,
  {\it A Four Dimensional Generalization of the Quantum Hall Effect}, 
  Science {\bf 294} (2001) 823
  [arXiv:cond-mat/0110572].
  
\bibitem{Chen:1999ab}
B.~Chen, H.~Itoyama and H.~Kihara,
{\it Nonabelian monopoles from matrices: Seeds of the spacetime
  structure}, 
Nucl.\ Phys.\ B {\bf 577}, 23 (2000)
[arXiv:hep-th/9909075].
 

\bibitem{Fabinger:2002bk}
  M.~Fabinger,
  {\it Higher-dimensional quantum Hall effect in string theory}, 
  JHEP {\bf 0205} (2002) 037
  [arXiv:hep-th/0201016].
  
\bibitem{Bernevig:2003yz}
B.~A.~Bernevig, J.~p.~Hu, N.~Toumbas and S.~C.~Zhang,
{\it The Eight Dimensional Quantum Hall Effect and the Octonions}, 
Phys.\ Rev.\ Lett.\  {\bf 91}, 236803 (2003)
[arXiv:cond-mat/0306045].

\bibitem{Meng:2003vj}
  G.~w.~Meng,
  {\it Geometric Construction of the Quantum Hall 
Effect in Higher Dimensions}, 
  J.\ Phys.\ A {\bf 36} (2003) 9415
  [arXiv:cond-mat/0306351]; 

\bibitem{Hasebe:2003gx}
  K.~Hasebe and Y.~Kimura,
  {\it Dimensional hierarchy in quantum Hall effects on fuzzy spheres}, 
  Phys.\ Lett.\ B {\bf 602} (2004) 255
  [arXiv:hep-th/0310274].

\bibitem{Meng:2004qe}
G.~w.~Meng,
{\it Dirac and Yang monopoles revisited}, 
arXiv:math-ph/0409051.
  
\bibitem{Grossman:1984pi}
  B.~Grossman, T.~W.~Kephart and J.~D.~Stasheff,
  {\it Solutions To Yang-Mills Field Equations In Eight-Dimensions 
And The Last
  Hopf Map}, 
  Commun.\ Math.\ Phys.\  {\bf 96} (1984) 431
  [Erratum-ibid.\  {\bf 100} (1985) 311].
  
\bibitem{Tchrakian:1978sf}
  D.~H.~Tchrakian,
  {\it N-Dimensional Instantons And Monopoles}, 
  J.\ Math.\ Phys.\  {\bf 21}, 166 (1980).

\bibitem{Saclioglu:1986qn}
  C.~Saclioglu,
  {\it Scale Invariant Gauge Theories And Selfduality In Higher Dimensions}, 
  Nucl.\ Phys.\ B {\bf 277}, 487 (1986).

\bibitem{Polchinski:2005bg}
  J.~Polchinski,
 {\it Open heterotic strings}, 
  arXiv:hep-th/0510033.


\bibitem{Randjbar-Daemi:1983qa}
  S.~Randjbar-Daemi, A.~Salam and J.~A.~Strathdee,
  {\it Instanton Induced Compactification And Fermion Chirality}, 
  Phys.\ Lett.\ B {\bf 132} (1983) 56.

\bibitem{Gibbons:1993sv}
  G.~W.~Gibbons and P.~K.~Townsend,
  {\it Vacuum interpolation in supergravity via super p-branes}, 
  Phys.\ Rev.\ Lett.\  {\bf 71} (1993) 3754
  [arXiv:hep-th/9307049].


\bibitem{Bartnik:1988am}
R.~Bartnik and J.~Mckinnon,
{\it Particle - Like Solutions Of The 
Einstein Yang-Mills Equations}, 
Phys.\ Rev.\ Lett.\  {\bf 61}, 141 (1988).


\bibitem{Radu:2005mj}
E.~Radu and D.~H.~Tchrakian,
{\it No hair conjecture, nonabelian hierarchies and 
anti-de Sitter spacetime}, 
Phys.\ Rev.\ D {\bf 73}, 024006 (2006)
[arXiv:gr-qc/0508033].

\bibitem{Cvetic:2003jy}
  M.~Cvetic, G.~W.~Gibbons, H.~Lu and C.~N.~Pope,
  {\it Consistent group and coset reductions of the bosonic string}, 
  Class.\ Quant.\ Grav.\  {\bf 20} (2003) 5161
  [arXiv:hep-th/0306043].


\bibitem{Fairlie:1984mp}
D.~B.~Fairlie and J.~Nuyts,
{\it Spherically Symmetric Solutions Of Gauge Theories In
  Eight-Dimensions}, 
J.\ Phys.\ A {\bf 17}, 2867 (1984).

\bibitem{Fubini:1985jm}
  S.~Fubini and H.~Nicolai,
  {\it The Octonionic Instanton}, 
  Phys.\ Lett.\ B {\bf 155} (1985) 369.

\bibitem{Gibbons:2005vp}
G.~W.~Gibbons, M.~J.~Perry and C.~N.~Pope,
{\it Bulk/boundary thermodynamic equivalence, and the Bekenstein and
cosmic-censorship bounds for rotating charged AdS black holes}, 
Phys.\ Rev.\ D {\bf 72}, 084028 (2005)
[arXiv:hep-th/0506233].

\bibitem{Okuyama:2002mh}
N.~Okuyama and K.~i.~Maeda,
{\it Five-dimensional black hole and particle solution 
with non-Abelian gauge field}, 
Phys.\ Rev.\ D {\bf 67}, 104012 (2003)
[arXiv:gr-qc/0212022].

\bibitem{Hosoya:1989zn}
A.~Hosoya and W.~Ogura,
{\it Wormhole Instanton Solution In The Einstein Yang-Mills System}, 
Phys.\ Lett.\ B {\bf 225}, 117 (1989).

\bibitem{Macfarlane:2002hr}
A.~J.~Macfarlane,
{\it The sphere S(6) viewed as a G(2)/SU(3) coset space}, 
Int.\ J.\ Mod.\ Phys.\ A {\bf 17}, 2595 (2002).

\bibitem{Ivanova:1993nu}
  T.~A.~Ivanova,
  {\it Octonions, selfduality and strings}, 
  Phys.\ Lett.\ B {\bf 315}, 277 (1993).

\bibitem{Gunaydin:1995ku}
  M.~G\"unaydin and H.~Nicolai,
  {\it Seven-dimensional octonionic Yang-Mills 
instanton and its extension to an
  heterotic string soliton}, 
  Phys.\ Lett.\ B {\bf 351} (1995) 169
  [Addendum-ibid.\ B {\bf 376} (1996) 329]
  [arXiv:hep-th/9502009].



 \end{thebibliography}
\end{document}